\documentclass[twocolumn,letterpaper]{IEEEAerospaceCLS}  

\usepackage[]{graphicx}    
\usepackage{chemformula}   
\usepackage{isotope}       
\newcommand{\ignore}[1]{}  

\begin{document}
\title{Interplanetary Rapid Transit Missions from Earth to Mars using Directed Laser Energy Driven Light Sails}

\author{%
Madhukarthik Mohanalingam\\
School of Aerospace Engineering\\
Georgia Institute of Technology\\
Atlanta, GA 30332\\
mmohanalingam3@gatech.edu	
\and 
Christopher E. Carr\\
School of Aerospace Engineering \&\\
School of Earth and Atm. Sciences\\
Georgia Institute of Technology\\
Atlanta, GA 30312\\
cecarr@gatech.edu
\thanks{\footnotesize 978-1-6654-9032-0/23/$\$31.00$ \copyright2023 IEEE}              
}

\maketitle

\thispagestyle{plain}
\pagestyle{plain}

\maketitle

\thispagestyle{plain}
\pagestyle{plain}

\begin{abstract}
Interest in the exploration of, and the establishment of a human settlement, on Mars is rapidly growing. To accomplish this, rapid transit will be required to bring important supplies and cargo. Current missions to Mars take at least 150 days, which would be too long in case of emergencies or urgent needs. Therefore, we propose the use of a cutting-edge technology that could allow transit times as short as 20 days: laser energy driven light sails. This propulsion method uses a ground-based laser array to propel a small, lightweight spacecraft attached to a light sail to very high speeds, enabling missions that are much faster than current missions. By utilizing a MATLAB model and a laser propulsion computational tool, we visualize and determine the optimal trajectories and departure windows for these missions. We discuss these trajectories and show that these missions are possible during specific launch windows within a 27-month time period between 2030-2032, but have practical challenges within this period as well. During solar conjunction, such rapid transit missions are limited due to the proximity of the sun, but rapid transits are possible during all orbital phases when the transit time requirement is relaxed. Laser arrays capable of generating up to 13 GW are necessary to enable 20-day missions with a 5 kg spacecraft, capable of carrying valuable lightweight cargo to astronauts, near conjunction, but only 0.55 GW is required around opposition. Required spacecraft velocities always exceed the solar system escape velocity and the trajectories are hyperbolic. A significant challenge for future work involves mechanisms and processes for deceleration and entry, descent, and landing. A ground-based laser array on Mars could address some aspects of this challenge, but orbital geometry limits deceleration potential, implying that payloads would need to be robust to large deceleration and impact g-loads. These 20-day missions to Mars can serve as precursors to more complex, distant missions. Spacecraft mass capabilities can be increased while also decreasing transit times by optimizing the laser array and light sail properties. Multiple spacecraft may also be launched and boosted simultaneously to carry more payload and decrease costs. This work is intended to serve as a proof of concept that lightweight payloads can be transported via such missions. Technologies enabling rapid transit missions can be developed in the next several decades and be applied to deep space missions to other celestial bodies and journeys to interstellar space.

\end{abstract}

\tableofcontents

\section{Introduction}
For centuries, humanity has looked up at our enigmatic neighboring red planet with a sense of awe and curiosity. As space exploration progressed, Mars quickly became a key destination for robotic exploration. In recent years, there has been a tremendous amount of interest in establishing a human presence on Mars. However, this requires the development of a settlement capable of sustaining humans for long periods of time. The materials required for the maturation of the Martian infrastructure must either be delivered to the planet or based on resources derived from Mars itself. A combination of the former and the latter methods should be used to facilitate the settlement process. While some resources can be acquired in situ, the infrastructure required to produce complex goods (e.g., computer chips) will not be available. 

The delivery of necessary supplies in a timely manner would be critical for astronauts to survive and to prosper. However, current propulsion technologies limit the transit time between Earth and Mars to a minimum of 150 days, and an average travel time between 150-300 days. Today’s propulsion technology consists of solid fueled, liquid fueled, and electric propulsion. Currently, the most common and efficient propulsion method to reach Mars is chemical propulsion. However, we seek to substantially decrease the current transit times to enable rapid transit missions to Mars. In order to achieve this goal, the maturation of more propulsion method concepts is necessary. Many concepts have been proposed for fast interstellar missions, and these concepts include antimatter annihilation, nuclear fusion, and laser driven light sails, among others \cite{Litchford2020}. All of these methods are expected to enable spacecraft to travel at extremely fast speeds and would allow probes to reach interstellar destinations, such as Alpha Centauri, within a human’s lifetime. Of these concepts, laser driven light sails may be a technology that can be effectively applied to rapid Earth-Mars missions because they have the benefit of working today \cite{PARKIN2018370}. Light sails already exist in the form of solar sails and are a proven technology. High power lasers are rapidly being developed; the Extreme Light Infrastructure for Nuclear Physics (ELI-NP) laboratory has the most powerful lasers, capable of providing a maximum power of 10 PW ($10^{7}$ GW) \cite{Tanaka2020}. Therefore, current lasers are capable of providing far more power than 20-day Earth-Mars transit missions would require, if admittedly only for short periods. The development and application of this propulsion technology for Earth-Mars rapid transit missions can also serve as a proof of concept and precursor mission for various advanced missions, such as deep space, interstellar, planetary defense, and extrasolar object interception missions.

Light sails are a type of propulsion system that uses radiation pressure exerted by sunlight or high energy laser beams on large mirrors. Solar sails are a tested technology that utilize radiation pressure exerted by the sun as means of propulsion. However, laser driven light sails would replace the sun with a large, ground-based array of moderately powerful lasers as the source of radiation pressure. Laser driven light sails require an array of lasers optimally located in a dry, high-altitude location on Earth. The array of lasers would be combined to form a single beam capable of providing power on the scale of gigawatts.
	
The laser beams would be directed onto a large, lightweight, and highly reflective light sail, which would be attached to the spacecraft with a payload \cite{PARKIN2018370}. The same technology concept described above was proposed by Breakthrough Starshot for an interstellar mission to Alpha Centauri \cite{PARKIN2018370}, but it can be scaled down for a rapid transit mission to Mars. Here we evaluate the possibility of a 20-day rapid transit mission between Earth and Mars by using laser driven light sails with extensions to outer solar system and more advanced missions.

\section{Methods}

\begin{figure*} 
    \centering
    \includegraphics[width=7in]{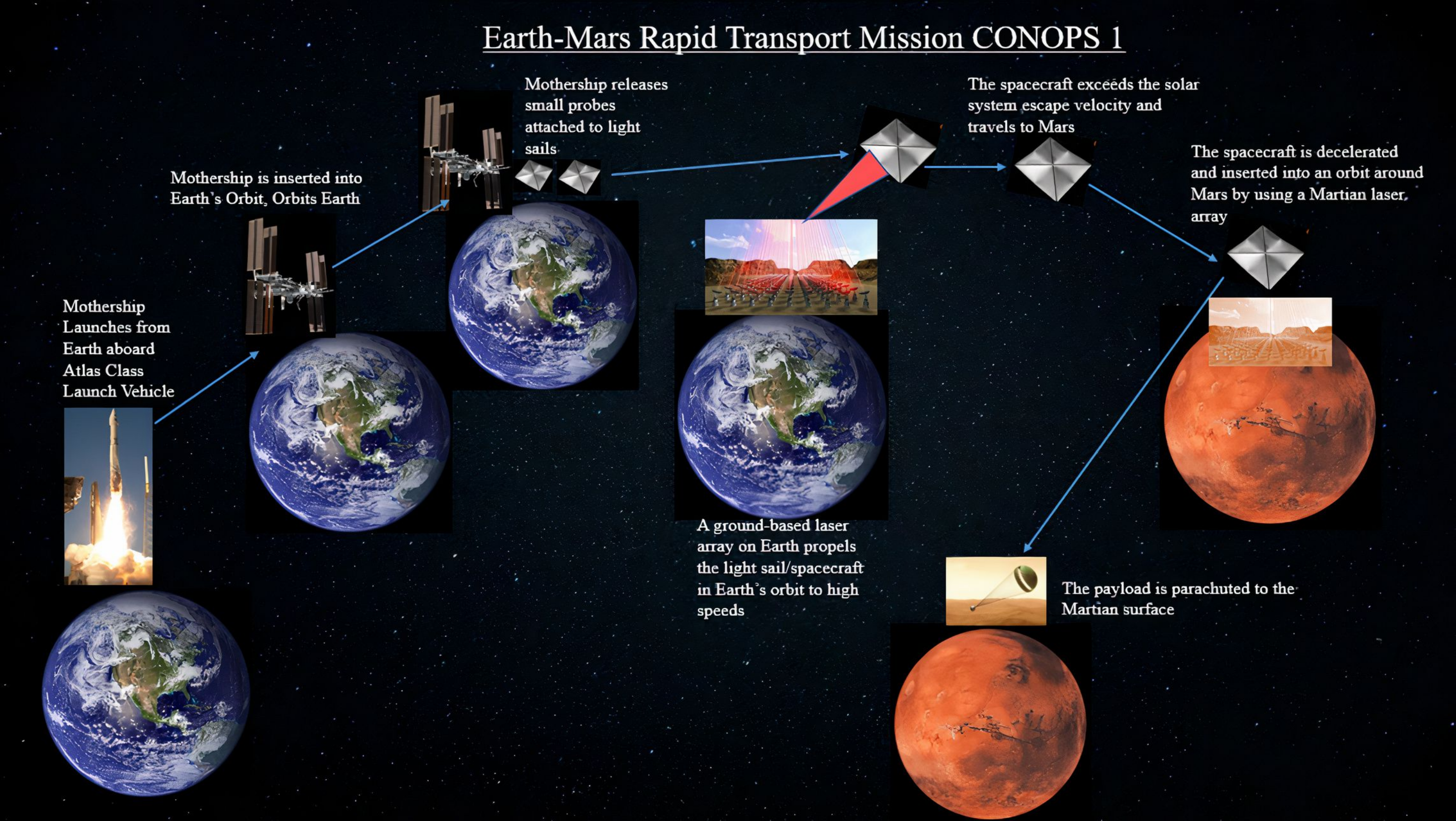}
    \caption{\bf{\label{fig:CONOPS}Concept of Operations (CONOPS) for Earth-Mars Rapid Transit}}
\end{figure*}

A 20-day mission time frame was selected as a nominal target in order to explore the potential for an order of magnitude improvement compared to low-energy transfers based on chemical propulsion. Next, trajectory modeling, velocity ($\Delta v$) computations, and propulsion system and spacecraft parameter calculations were completed. Prior to this, however, several concepts of operations (CONOPS) were developed for the mission. On the basis of the most efficient and practical mission operations, a nominal CONOPS was selected (Figure ~\ref{fig:CONOPS}).
The intended phases are:            \begin{itemize}
    \item Launch and Orbit Insertion: The spacecraft launches from Earth’s surface aboard a mothership and is inserted into a low Earth parking orbit by the launch vehicle.\\ 
    \item Sail Deployment: The light sail of the spacecraft is deployed as the spacecraft is coasting in the low Earth parking orbit after being released from the mothership.\\
    \item Boost Phase: A ground-based array of lasers are all combined and beamed directly at the light sail. This boosts the spacecraft to high velocities and puts it on its trajectory to Mars. Ideally, this phase only lasts a few hours.\\
    \item Cruise Phase: The spacecraft cruises through space on a hyperbolic trajectory to Mars; it will not be bound by the solar system because the velocity will be higher than the solar system escape velocity.\\ 
    \item Deceleration Phase: The spacecraft is decelerated as it reaches Mars, via direct entry (our assumption in this work) or possibly via deceleration by an alternate ground-based laser array on Mars (not modeled in this work). Alternative methods may include deceleration with aerocapture methods or magnetic and/or electric sails.\\
    \item Entry, descent, and landing phase: The payload of the spacecraft arrives on the Martian surface (not modeled in this work).\\ 
\end{itemize}

Next, the trajectories were simulated using the MATLAB SPICE toolkit provided by the NASA Jet Propulsion Laboratory. The SPICE toolkit is an astrodynamics modeling tool that provides a library and application program to facilitate the modeling of the orbits of planets and spacecraft \cite{ACTON199665}\cite{ACTON20189}.

By implementing the functions provided by the toolkit with additional code that was developed in-house, critical calculations and plots were generated. The developed code applied the Lancaster \& Blanchard solution, with improvements by Gooding, to solve Lambert’s (Gauss') problem, which allows the computation of the departure and arrival velocity of a spacecraft traveling between two different astronomical bodies. The overall MATLAB program to solve Lambert's problem featured two different approaches. Of the two approaches, the more robust method was the  Lancaster \& Blanchard solution, with improvements by Gooding \cite{Blanchard1969}\cite{Gooding1990}\cite{rodyo}. Therefore, this method was selected to be incorporated into our developed MATLAB model for computations due to our desire of solving Lambert's problem for many diverse cases. Additionally, another personally developed MATLAB model employed the universal variables method to solve Lambert's problem; this model was used to verify the results. By utilizing the MATLAB models, $V_{infinity}$,  departure velocity, and arrival velocity, were all computed for a series of Earth departure and Mars arrival dates. The feasibility of the trajectories was then assessed by examining the required optical power and the trajectory's proximity to the Sun.

Trajectories and corresponding calculations were determined for a variety of dates over the course of a 27-month cycle. Mars oppositions occur when the Earth passes between the Sun and Mars, making opposition the time around when Mars is closest to the Earth. The Mars oppositions occur around every 26 months, and so one specific 27-month cycle was chosen to be the time frame for the potential trajectory. A variety of dates were chosen for the Earth departure dates and Mars arrival dates between April 25, 2030, and August 10, 2032. Within this time frame, Mars is at opposition on May 4, 2031. Solar conjunction occurs twice within this time period – once on May 25, 2030, and once on July 11, 2032. Solar conjunction is when the sun passes between the Earth and Mars, making conjunction the time around when Mars is at one of the farthest distances away from the Earth. Therefore, a wide variety of planetary positions and velocities were tested by choosing this specific range of dates. The MATLAB simulation was validated to be correct by modeling the trajectory of the Mars Reconnaissance Orbiter. 

Lubin developed the basis for estimating the physical and performance characteristics of the light sail, laser array, and the spacecraft itself in his paper \cite{Lubin2016}. The basis was implemented by the University of California, Santa Barbara, as a laser propulsion computational tool\footnote[1]{https://rjdlee.com/projects/calculator/}. Running the tool requires various inputs, including payload mass, sail physical properties, and optical power. The outputs of the tool included the sail mass, boost time, and spacecraft velocity. The departure velocity values initially computed from the MATLAB simulation were assumed to be the speed at $L_{0}$ output of the laser propulsion tool. This is the speed of the spacecraft at the point $L_{0}$, which is the location where the boost phase of the mission ends, meaning that the lasers stop propelling the spacecraft (distance where laser spot size equals the size of the light sail). The optical power and target distance inputs of the tool were varied to achieve the target departure velocity as the speed at $L_{0}$ output of the tool. All other inputs were kept constant for each trajectory. Table ~\ref{tab:inputs} shows relevant inputs of the tool and their values for an example case. The example case has an optical power of 1 GW and a target distance of around $9.12 \times 10^{-13}$ light years, which is the Earth-Mars distance when Mars is at opposition, and equals $8.63 \times 10^{7}$ km.



\begin{table}[]
\renewcommand{\arraystretch}{1.3}
\caption{\label{tab:inputs}Inputs of the Laser Propulsion Tool with 1 GW of Power when Mars is at Opposition.}
\centering
\begin{tabular}{|l|l|l|}
\hline
\textbf{Inputs} & \textbf{Value} & \textbf{Unit} \\
\hline
Payload & 5 & kg \\
\hline
Sail Side Length & 19.86 & m \\
\hline
Sail Thickness & 4 & $\mu$m \\
\hline
Sail Density & 3.17 & g/cm${^3}$ \\
\hline
Sail Reflection Efficiency (0   - 1) & 0.9 & {} \\
\hline
Sail Light Absorptivity (0 - 1) & 0.1 & {} \\
\hline
Sail Front Emissivity (0 - 1) & 0.5 & {} \\
\hline
Sail Back Emissivity (0 - 1) & 0.5 & {} \\
\hline
Laser Array Side Length & 500 & m \\
\hline
Total Optical Power & 1 & GW \\
\hline
Beam Efficiency (0 - 1) & 0.7 & {} \\
\hline
Wavelength & 1060 & nm \\
\hline
Electrical Efficiency (0 - 1) & 0.5 & {} \\
\hline
Electrical Energy Cost & 0.1 & \$/kW-hr \\
\hline
Energy Storage Cost & 0.1 & \$/W-hr \\
\hline
Target Distance & $9.12 \times 10^{-13}$ & ly \\
\hline
\end{tabular}
\end{table}

\begin{table}[]
\renewcommand{\arraystretch}{1.3}
\caption{\label{tab:outputs}Outputs of the Laser Propulsion Tool with 1 GW of Power when Mars is at Opposition.}
\centering
\begin{tabular}{|l|l|l|} \hline
\textbf{Outputs} & \textbf{Value} & \textbf{Unit} \\ \hline
Sail Mass & $5.00 \times 10^{3}$ & g \\ \hline
Total Mass & $1.00 \times 10^{4}$ & g \\ \hline
Areal Density & 12.7 & g/m$^{2}$ \\ \hline
Laser Power in Main Beam & 0.7 & GW \\ \hline
Flux on Sail & 0.00178 & GW/m$^{2}$ \\ \hline
Total Force on Sail & 4.23 & N \\\hline
Peak Sail Pressure & 0.0107 & Pa \\\hline
Peak Acceleration & 0.423 & m/s$^{2}$ \\ \hline
Distance at End of Boost $L_{0}$ & 0.0313 & au \\ \hline
Time to $L_{0}$ & 1.72 & d \\ \hline
Speed at $L_{0}$ & 62.9 & km/s \\ \hline
Kinetic Energy at $L_{0}$ & 0.0055 & GW$\times$hr \\ \hline
\begin{tabular}[c]{@{}c@{}}Laser Energy in \\ Main Beam at   $L_{0}$\end{tabular} & 28.9 & GW$\times$hr \\ \hline
Electrical Energy at $L_{0}$ & 82.7 & GW$\times$hr \\ \hline
\begin{tabular}[c]{@{}c@{}}Launch Efficiency at $L_{0}$ \\ (KE /   Electrical Energy)\end{tabular} & 0.00665 & \% \\ \hline
Electrical Energy Cost at $L_{0}$ & $8.27 \times 10^{6}$ & \$ \\ \hline
Energy Storage Cost at $L_{0}$ & $8.27 \times 10^{9}$ & \$ \\ \hline
Limiting Speed & 89 & km/s \\ \hline
Time to Target at Limiting Speed & 0.143 & yr \\ \hline
Max Flux Absorbed by Sail & $1.78 \times 10^{-5}$ & GW/m$^{2}$ \\ \hline
Max Power Absorbed by Sail & 0.007 & GW \\ \hline
Sail Equilibrium Temperature & 748 & K \\\hline
\end{tabular}
\end{table}

Again, note that all of the inputs of Table ~\ref{tab:inputs} were kept the same except for the target distance and optical power. The amount of optical power used directly affected the velocities that the spacecraft can reach. It was assumed that a square laser array with a side length of 500 meters (0.311 miles) was used. Therefore, the total area of the laser array would be 250,000 m$^{2}$ ($2.69 \times 10^6$ ft$^{2}$). Additionally, the sail’s physical parameters were determined by assuming that the sail is a square in shape and is made of silicon nitride, which is a strong candidate for the sail material \cite{Tung2021}. Some inputs involving communications are not included because they are not relevant for computing the specific outputs of interest. The only input that is set to be affected by other inputs and so is not manually selected in the computational tool is the optimal sail side length, which changes based on the payload, sail thickness, and sail density. The tool computes this optimal sail side length by assuming that the sail mass equals the payload mass because this is outlined as an optimal case for maximizing velocity in Lubin's paper \cite{Lubin2016}. Therefore, the sail mass was set to be equivalent to the payload mass. If we assume the payload, sail thickness, and sail density to be constant at 5 kg, 4 $\mu$m, and 3.17 g/cm$^{3}$, the sail side length is also set to remain constant. If these sail properties are kept constant, the sail size can be fixed and replicated for alternate missions. Table ~\ref{tab:outputs} shows all the outputs of the laser propulsion tool for the specific input case shown in Table ~\ref{tab:inputs}.

\section{Results}
The MATLAB simulation and laser propulsion computational tool were used in conjunction to retrieve the ideal departure windows for 20-day missions to Mars over a 27-month cycle, the optimal trajectories for these 20-day missions to Mars, and the overall feasibility of such trajectories. The MATLAB simulation provided the optimal Earth-Mars trajectory, including the required departure and arrival velocities of the spacecraft. Additionally, the simulation generated visualizations of the optimal spacecraft trajectories for various Earth departure and Mars arrival dates. As stated earlier, the laser propulsion tool was utilized, and the input optical power of the tool was optimized to ensure that the $V_{0}$ (velocity at $L_{0}$) output was equal to the departure velocity. Therefore, the optimal optical power was computed by using the departure velocity value determined from the simulation. The time to $L_{0}$ values were also computed with the laser propulsion tool since these values signify the boost phase duration of the mission. The results were used to determine the trajectories and specific departure date ranges that actually allowed 20-day missions to Mars over the April 25, 2030 to July 11, 2032 time period. Table ~\ref{tab:trajchar} shows the 20-day Earth-Mars trajectory characteristics for some example departure dates. The key characteristics included are the Earth-Mars distance given the launch date and 20-day transit time, spacecraft departure velocity from Earth, and the optical power required to achieve the mission. Table ~\ref{tab:proximity} shows the different 20-day transit departure dates along with the closest position of the spacecraft to the Sun's center during its trajectory. Table ~\ref{tab:TLO} displays the time to reach $L_{0}$, which the boost phase duration, and the optical power required for 20-day transit on a given departure date. The time to reach $L_{0}$ is relevant because it is the point at which the spacecraft's boost phase ends because the laser spot equals the entire sail size.

\begin{table}[]
\renewcommand{\arraystretch}{1.3}
\caption{\label{tab:trajchar}Trajectory Characteristics for 20-day Earth-Mars Transit.}
\centering
\begin{tabular}{|l|l|l|l|}
\hline
\textbf{\begin{tabular}[c]{@{}l@{}}Departure\\      Date\end{tabular}} & \textbf{\begin{tabular}[c]{@{}l@{}}Earth-Mars \\ Distance \\ (km)\end{tabular}} & \textbf{\begin{tabular}[c]{@{}l@{}}Departure \\ Velocity \\ (km/s)\end{tabular}} & \textbf{\begin{tabular}[c]{@{}l@{}}Optical \\ Power \\ (GW)\end{tabular}} \\
\hline
4/25/2030 & $3.66 \times 10^{8}$ & 209.130 & 11.088 \\
\hline
5/25/2030 & $3.79 \times 10^{8}$ & 215.950 & 11.732 \\
\hline
6/25/2030 & $3.87 \times 10^{8}$ & 220.230 & 12.283 \\
\hline
9/5/2030 & $3.74 \times 10^{8}$ & 215.510 & 11.732 \\
\hline
12/1/2030 & $2.97 \times 10^{8}$ & 171.900 & 7.430 \\
\hline
2/1/2031 & $2.08 \times 10^{8}$ & 121.380 & 3.730 \\
\hline
4/4/2031 & $1.18 \times 10^{8}$ & 70.670 & 1.261 \\
\hline
5/4/2031 & $8.63 \times 10^{7}$ & 53.360 & 0.719 \\
\hline
6/5/2031 & $7.39 \times 10^{7}$ & 46.470 & 0.546 \\
\hline
9/5/2031 & $1.32 \times 10^{8}$ & 77.780 & 1.528 \\
\hline
12/1/2031 & $2.10 \times 10^{8}$ & 122.210 & 3.791 \\
\hline
3/1/2032 & $3.03 \times 10^{8}$ & 174.690 & 7.693 \\
\hline
6/11/2032 & $3.84 \times 10^{8}$ & 219.670 & 12.171 \\
\hline
7/11/2032 & $3.95 \times 10^{8}$ & 225.320 & 12.846 \\
\hline
8/10/2032 & $3.98 \times 10^{8}$ & 227.100 & 13.075 \\
\hline
\end{tabular}
\end{table}

\begin{table}[]
\renewcommand{\arraystretch}{1.3}
\caption{\label{tab:proximity}Proximity of Trajectory to the Sun for 20-day Earth-Mars Transit.}
\centering
\begin{tabular}{|l|l|l|}
\hline
\textbf{\begin{tabular}[c]{@{}l@{}}Departure\\      Date\end{tabular}} & \textbf{\begin{tabular}[c]{@{}l@{}}Closest S/C \\ Position to \\ Sun's Center \\ (km)\end{tabular}} & \textbf{Notes} \\
\hline
4/25/2030 & $4.54 \times 10^{7}$ &  \\
\hline
5/25/2030 & $3.03 \times 10^{7}$ & Conjunction \\
\hline
6/25/2030 & $2.35 \times 10^{7}$ &  \\
\hline
9/5/2030 & $6.85 \times 10^{7}$ &  \\
\hline
12/1/2030 & $1.26 \times 10^{8}$ &  \\
\hline
2/1/2031 & $1.47 \times 10^{8}$ &  \\
\hline
4/4/2031 & $1.50 \times 10^{8}$ &  \\
\hline
5/4/2031 & $1.51 \times 10^{8}$ & Opposition \\
\hline
6/5/2031 & $1.52 \times 10^{8}$ &  \\
\hline
9/5/2031 & $1.51 \times 10^{8}$ &  \\
\hline
12/1/2031 & $1.39 \times 10^{8}$ &  \\
\hline
3/1/2032 & $1.05 \times 10^{8}$ &  \\
\hline
6/11/2032 & $4.65 \times 10^{7}$ &  \\
\hline
7/11/2032 & $2.88 \times 10^{7}$ & Conjunction \\
\hline
8/10/2032 & $2.57 \times 10^{7}$ & \\
\hline
\end{tabular}
\end{table}

\begin{table}[]
\renewcommand{\arraystretch}{1.3}
\caption{\label{tab:TLO}Time to L$_{0}$ and Power Required for 20-day Earth-Mars Transit.}
\centering
\begin{tabular}{|l|l|l|}
\hline
\textbf{\begin{tabular}[c]{@{}l@{}}Departure\\      Date\end{tabular}} & \textbf{\begin{tabular}[c]{@{}l@{}}Time to\\ L$_{0}$ \\ (days)\end{tabular}} & \textbf{\begin{tabular}[c]{@{}l@{}}Optical \\ Power \\ (GW)\end{tabular}} \\
\hline
4/25/2030 & 0.517 & 11.088 \\
\hline
5/25/2030 & 0.503 & 11.732 \\
\hline
6/25/2030 & 0.492 & 12.283 \\
\hline
9/5/2030 & 0.503 & 11.732 \\
\hline
12/1/2030 & 0.632 & 7.430 \\
\hline
2/1/2031 & 0.892 & 3.730 \\
\hline
4/4/2031 & 1.53 & 1.261 \\
\hline
5/4/2031 & 2.03 & 0.719 \\
\hline
6/5/2031 & 2.33 & 0.546 \\
\hline
9/5/2031 & 1.39 & 1.528 \\
\hline
12/1/2031 & 0.885 & 3.791 \\
\hline
3/1/2032 & 0.621 & 7.693 \\
\hline
6/11/2032 & 0.494 & 12.171 \\
\hline
7/11/2032 & 0.481 & 12.846 \\
\hline
8/10/2032 & 0.477 & 13.075 \\
\hline
\end{tabular}
\end{table}

By analyzing Table ~\ref{tab:trajchar}, one can observe that the required spacecraft velocities to enable 20 day transit times increase as the distance between Earth and Mars increases. The optical power required to propel the spacecraft to Mars in 20 days increases as the required velocities for the spacecraft increase. The largest departure velocity required for the 20-day mission, 227.10 km/s, occurs for a departure date on August 10, 2032, 30 days after solar conjunction. This is the date when Earth and Mars are the farthest apart from each other as a result of solar conjunction, when the Sun passes between Earth and Mars. Additionally, the smallest departure velocity required for the 20-day mission, 46.47 km/s, occurs for a departure date on June 5, 2031, 32 days after Mars is at opposition. This is the date when Earth and Mars are the closest to each other as a result of Martian opposition, when the Earth passes between the Sun and Mars. All calculations were computed and trajectories generated relative to the J2000 reference frame. Theoretically, it is possible to enable 20-day missions near the time of solar conjunction. However, a major limitation is how close the spacecraft passes to the sun on its trajectory. If the spacecraft trajectory requires it to travel in very close proximity to the Sun, the spacecraft may be affected by heat and solar activity. 

\begin{figure*} 
    \centering
    \includegraphics[width=5in]{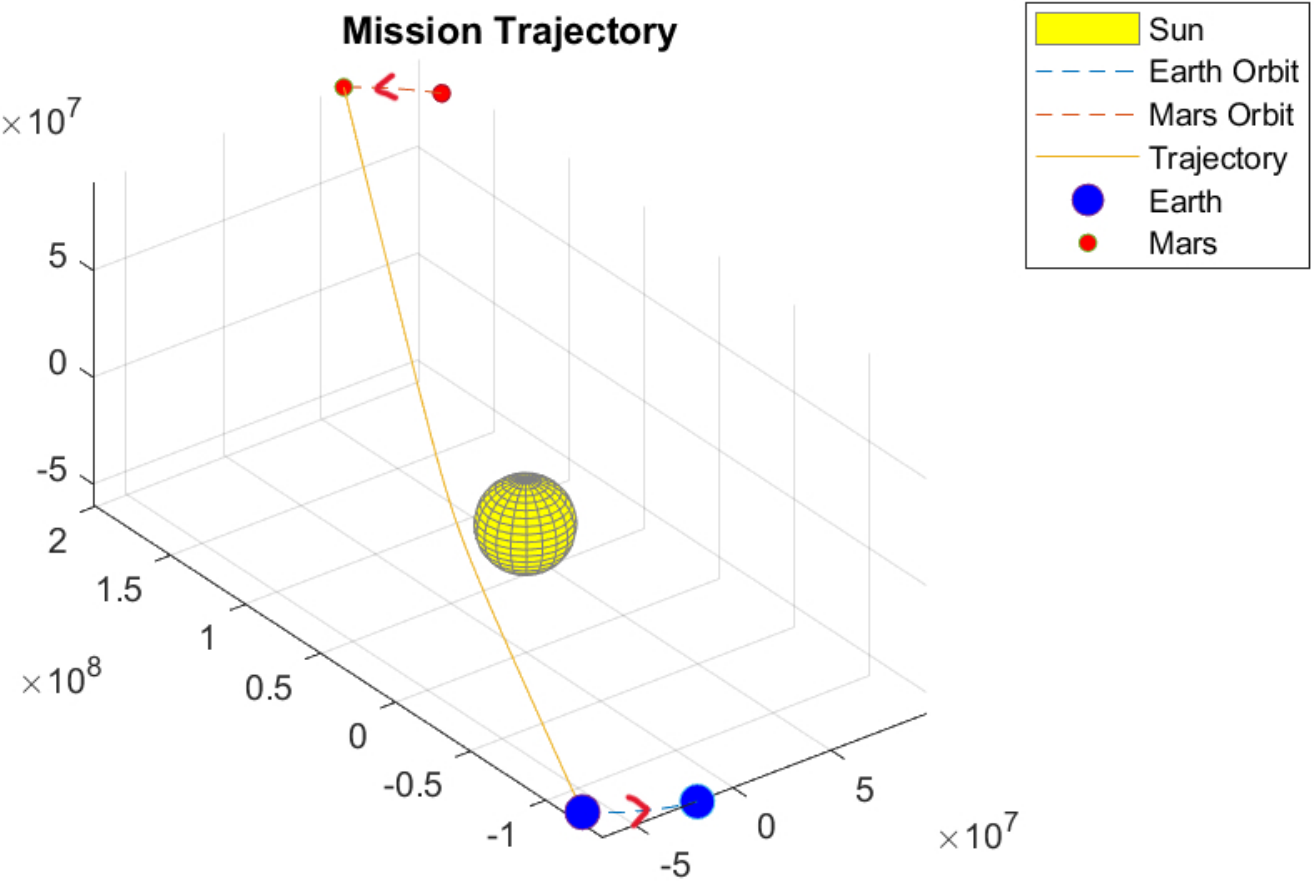}
    \caption{\bf{\label{fig:Conjunctionv2}20-Day Trajectory with Departure on Solar Conjunction}}
\end{figure*}

\begin{figure*} 
    \centering
    \includegraphics[width=5in]{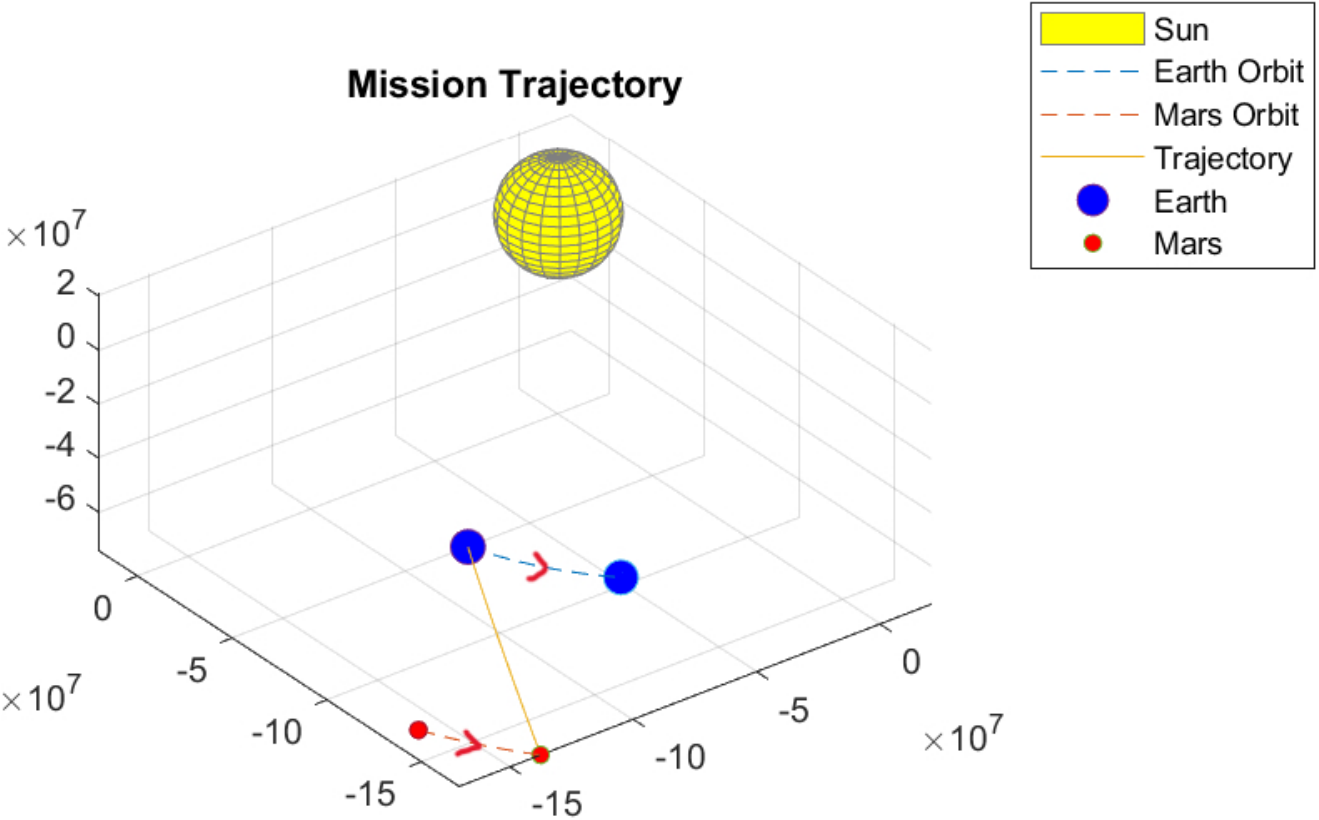}
    \caption{\bf{\label{fig: Oppositionv2}20-Day Trajectory with Departure on Martian Opposition}}
\end{figure*}

As a result, the feasibility of the 20-day missions would be reliant on the positions of Earth and Mars and the corresponding distance between them. Therefore, rapid transit missions should depart between these departure windows to ensure the most efficient missions. The 20-day trajectory for departure during solar conjunction on May 25, 2030 is shown in Figure ~\ref{fig:Conjunctionv2}. 
As a side note, a scale factor was applied to the Sun in order to emphasize its presence and location. A similar scale factor was applied to the Earth and Mars as well in the trajectory figures. Figure ~\ref{fig:Conjunctionv2} clearly shows that this trajectory will take the spacecraft to very close proximity to the Sun and also shows that Earth and Mars are very far apart. As a result, the feasibility of a 20-day mission is harder due to the Sun's close proximity and the larger amount of optical power required to enable the spacecraft to reach the high velocity necessary. Several trajectories with departure dates around conjunction were also generated, and all these trajectories showed that the optimal trajectory required the spacecraft to travel close to the sun at very high velocities requiring large amounts of optical power. Conversely, trajectories on and around Martian opposition drastically improved the feasibility of 20-day missions. The feasibility improved because lower departure velocities were required since the distance between Earth and Mars was smaller. As a result, a much smaller amount of optical power was necessary. Additionally, solar proximity was not an issue, meaning the Sun did not pose as a threat to the spacecraft. Additionally, it must be noted that the Sun's radiation pressure and gravitational pull may also interfere with the spacecraft trajectory. However, these effects were assumed to be negligible for the purposes of this project due to the high velocities of the spacecraft. The trajectory for a 20-day mission with a departure date on the date of Martian opposition, occurring on May 4, 2031, is shown in Figure ~\ref{fig: Oppositionv2}. 

Both trajectories shown in Figure ~\ref{fig:Conjunctionv2} and Figure ~\ref{fig: Oppositionv2} are direct, hyperbolic trajectories since the interplanetary trajectories all have orbital eccentricities of greater than 1. The spacecraft follows a hyperbolic interplanetary trajectory and is not bound to our solar system. This occurs since the required spacecraft departure velocities are higher than the solar system escape velocity of 42.1 km/s. Therefore, the spacecraft is not bound to the Sun during its transit to Mars. 

\section{Discussion}
The results in the previous section give insights regarding the optimal trajectory characteristics for 20-day rapid transit missions. A recent study evaluated 45-day missions to Mars by using laser-thermal propulsion \cite{DUPLAY2022143}. Although this method also used lasers, it did not use light sails and also added a thermal propulsion component. Improving upon 45-day transit times, proper 20-day trajectories are found to be possible during a majority of the time during a 27 month time period that also includes a month preceding and following the two dates of solar conjunction within this time frame. Martian opposition occurs within this time period as well, and the specific range of dates covered range from April 25, 2030 to August 10, 2032. In this work, we considered a 5 kg spacecraft mass. A key question to answer is whether this 5 kg mass provides any useful payload. The space industry has been moving towards the miniaturization of systems, with low mass CubeSats being used regularly. Therefore, lightweight cargo, such as computer chips, small medical items, mini cameras, and other miniature items, instruments, and appliances for personal and scientific use can be within the mass constraint. Despite the limitation of payload mass, valuable miniature cargo can still be transported to astronauts using this spacecraft concept. As mentioned earlier, this work is intended to serve as a proof of concept to demonstrate the basic capabilities of such missions, and can be scaled to allow additional payload to be carried by the spacecraft. In order to increase payload mass while also decreasing optical power requirements, the spacecraft sail and laser array parameters can be changed. The size of the laser array can be increased and laser efficiencies can be improved. A reduction in the sail mass also enables an increase in payload mass. The sail mass can be decreased by reducing sail thickness, sail side length, and sail density. However, there are limits regarding how small the sail's mass and size can be made due to sail size requirements. Ultimately, the sail material may be changed to a single or mixture of optimal materials that provides high reflectivity while also having low density and thickness. Conversely, decreasing payload mass decreases the required optical power. The lightweight spacecraft is comparable in size to a 2.5 U CubeSat due to its mass of 5 kg. The spacecraft's large light sail has a separate mass that is not included in this 5 kg. The spacecraft does not carry any fuel for its main propulsion system, so no mass is necessary for a propulsion system onboard the spacecraft other than the light sail itself. Light sails are beneficial since they theoretically have an infinite specific impulse as a result of not carrying any fuel. However, it would likely be beneficial to carry at least some fuel onboard as a back-up for redundancy.  It was found that decreasing the payload size increases the velocity that the spacecraft itself can reach, and this decreases the transit time of the spacecraft. Spacecraft velocity increases as optical power increases, resulting in the decrease in transit time. Additionally, a decrease in payload size decreases the size of the light sail itself as well. 

The trajectories are all found to be hyperbolic and are not bound by the solar system since the spacecraft departure velocities are all greater than the Sun's escape velocity of 42.1 km/s. The required velocity values range from 46.5 km/s to 227.1 km/s for departure dates within the selected time frame. Higher velocities require higher optical power values from the laser array, and these values range from 0.55 GW to 13 GW. Therefore, different amounts of power will need to be generated by the laser array based on the required velocities necessary for the mission, which is based on the spacecraft departure date itself since the positions of Earth, Mars, and Sun vary over the 839 day time frame considered. As a result, the laser array must be built with the capability of either shutting down a certain number of lasers based on the power need, ensuring that the lasers can provide a variety of amounts of power to meet the power requirements, or reducing the boost time if other limits are met. 

\subsection{Limitations and Future Development}
It is important to discuss the limitations of the mission concept and the areas requiring future research. Some limitations that must be addressed include the boost phase duration (Time to $L_{0}$), the Sun's proximity to the spacecraft during solar conjunction, and the deceleration phase of the spacecraft.   

A key factor that must be discussed is the boost phase duration, which is the Time to $L_{0}$ parameter that the laser propulsion tool outputs. This parameter varies based on the optical power input. As optical power input increases, the boost time decreases. This time represents how long it takes for the spacecraft to reach the required velocity to be put on the proper 20-day trajectory to Mars. Specifically, as calculated here, this time is when the laser boost phase ends because the laser spot equals the entire sail size \cite{Lubin2016}. To allow boost by a single laser array on the surface of Earth, this time must be minimized. It is important to limit the requirement to aim the lasers across a wide range of angles due to Earth's rotation and avoid low angles where laser light would pass through a lot of atmosphere. It is important to note that the spacecraft can only be boosted when the spacecraft is visible from the laser array on the Earth's surface. Therefore, the initial orbit of the spacecraft around Earth must be selected to minimize the boost phase duration. However, the boost times predicted range from .477 days to 2.33 days, which are substantially large periods of time that the lasers must continuously boost the spacecraft. Given large boost times, the lasers must have the ability to change their orientation significantly or multiple laser arrays would be required as Earth rotates in order to ensure that the spacecraft sail is boosted continuously. However, this can be challenging since the lasers must align properly with the spacecraft and account for the rotation of the Earth. If the lasers cannot be oriented properly for the entire time of the boost phase, another option is to periodically shut down the lasers during the boost phase itself. This would allow the lasers to only boost the spacecraft when they are in the proper orientations relative to the spacecraft. Higher laser power could compensate to achieve the required velocity in a shorter time period (e.g., before reaching $L_{0}$). In this case, laser power is the same regardless of how many light sails are accelerated (for small numbers of spacecraft that would fit within the beam) for boost phases to distances much less than $L_{0}$. Therefore, it may be beneficial to send several spacecraft at the same time to ensure that more payload can be taken since payload masses can be split, meaning less optical power and costs are necessary. Another option is to have several laser arrays strategically placed around Earth to continuously boost the spacecraft in orbit. Another MATLAB model was developed in-house to derive optical power requirements by constraining the boost phase duration. The inputs of the model were the boost phase duration constraint, target velocities, and spacecraft physical parameters. It was determined that constraining the boost phase duration resulted in higher optical power values for given target velocities. The distance of the boost phase also increased; this is the distance that the spacecraft must be boosted from its original orbit around Earth. The optical power values increased by substantial factors for the various target velocities, ranging from a factor of 2.5 to 13. Therefore, constraining boost phase duration requires the optical power to increase dramatically. 



\begin{figure*} 
    \centering
    \includegraphics[width=5in]{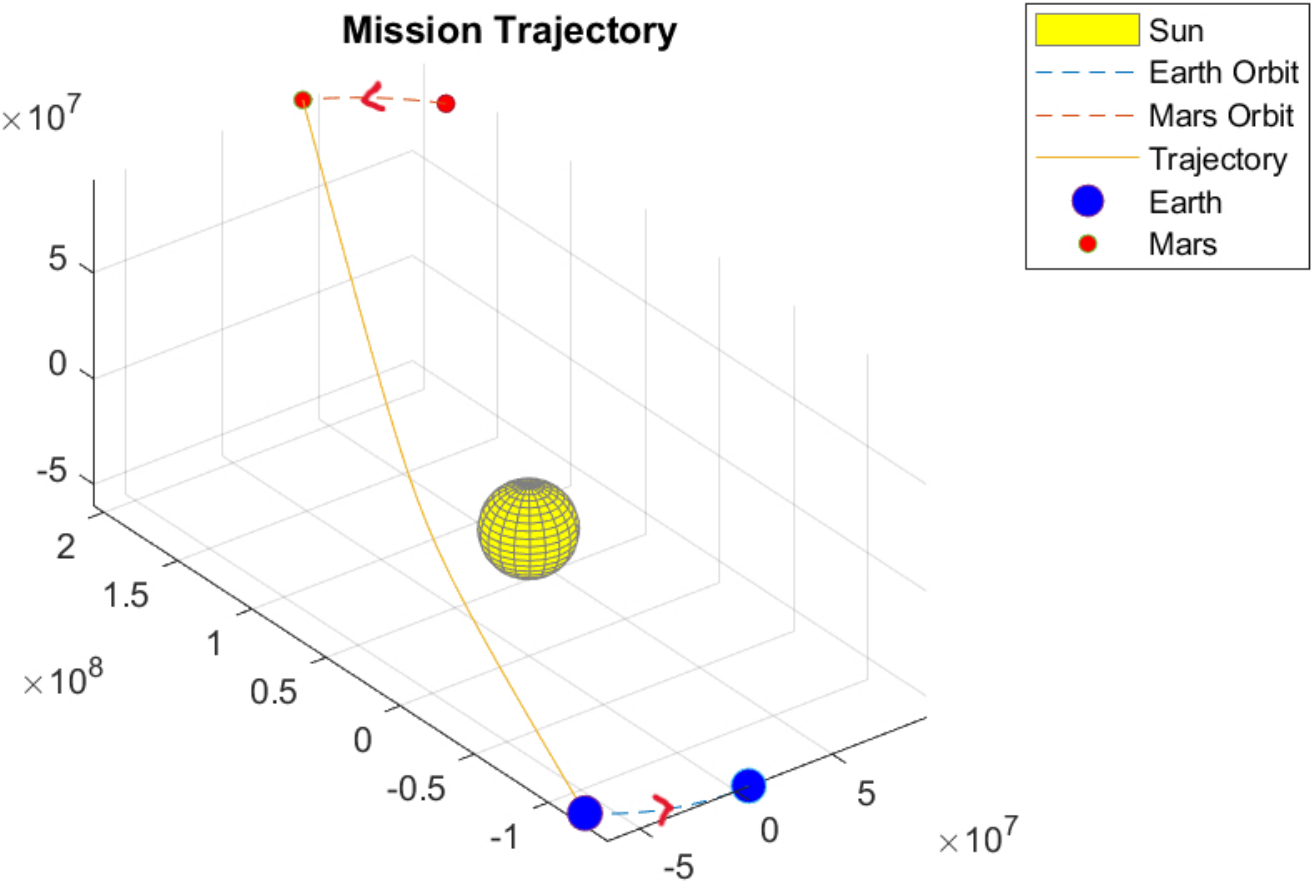}
    \caption{\bf{\label{fig:Conjunction2v2}30-Day Trajectory with Departure on Solar Conjunction}}
\end{figure*}

As mentioned in the previous section, 20-day Earth-Mars missions are theoretically possible during and around solar conjunction, but a limitation is the proximity of the sun to the spacecraft trajectory. If the spacecraft passes too close to the Sun, the function and integrity of the spacecraft and its components may be affected. Therefore, the spacecraft must have robust thermal protection systems to ensure that its functional capabilities and structural integrity are not compromised due to the immense heat it may be subjected to. During and around Martian opposition, the proximity of the Sun is not a significant issue for concern. Furthermore, the spacecraft will not be in close proximity to the Sun for the majority of time during the selected time period. Specifically, the time frame when the spacecraft is dangerously close to the Sun starts around a month before solar conjunction and ends around 2 months after solar conjunction. The closest approach to the Sun by a spacecraft will be achieved by the Parker Solar Probe, which is expected to reach 6.16 million km from the Sun's surface \cite{Venzmer2018}. The second closest spacecraft to the Sun is Helios 2, which came within 43.43 million km from the Sun's center \cite{Venzmer2018}. For this mission concept, the Helios 2 approach is used as a benchmark for the closest distance that the spacecraft can be from the Sun, since the Parker solar probe has advanced thermal protection systems that would add more mass and cost. A cheaper, contemporary thermal protection system similar to that of Helios 2 would work well for the rapid transit spacecraft. Longer transit times can help avoid solar proximity during conjunction and can decrease the spacecraft velocity for reducing optical power requirements. Therefore, increasing the transit time of 20-day missions around solar conjunction allows for improved efficiency and lowered risk. Table ~\ref{tab:longertime} shows longer transit times around conjunction that still allow for rapid transit missions compared to the current standard. It can be observed in Table ~\ref{tab:longertime} that the departure velocities and optical power requirements decrease substantially when the transit time is increased from 20 days. Figure ~\ref{fig:Conjunction2v2} shows a 30-day mission to Mars with a departure date on the date of solar conjunction on May 25, 2030. By comparing Figure ~\ref{fig:Conjunction2v2} and Figure ~\ref{fig:Conjunctionv2}, it is easy to observe that the 30-day trajectory of the spacecraft allows it to travel farther away from the Sun when compared to the 20-day trajectory. A trade-off between increased transit time and lower velocity and power requirements is important to consider for missions departing on and around solar conjunction. Overall, a proper thermal protection system that does not add an excessive amount of spacecraft mass would be ideal for the rapid transit spacecraft to ensure that solar proximity is not an issue during missions around solar conjunction. 

\begin{table}[] 
\renewcommand{\arraystretch}{1.3}
\centering
\caption{\label{tab:longertime}Trajectories with Longer Transit Times near Solar Conjunction}
\begin{tabular}{|l|l|l|l|} \hline 
{\textbf{\begin{tabular}[c]{@{}c@{}}Departure\\      Date\end{tabular}}} & {\textbf{\begin{tabular}[c]{@{}c@{}}Transit \\Time\\      (Days)\end{tabular}}} & {\textbf{\begin{tabular}[c]{@{}c@{}}Departure\\       Velocity (km/s)\end{tabular}}} & {\textbf{\begin{tabular}[c]{@{}c@{}}Optical \\      Power (GW)\end{tabular}}} \\\hline
5/25/2030 & 30 & 142.85 & 5.130 \\ \hline
6/25/2030 & 41 & 106.63 & 2.866 \\ \hline
7/11/2032 & 32 & 139.47 & 4.916 \\ \hline
8/10/2032 & 43 & 104.44 & 2.759 \\ \hline
\end{tabular}
\end{table}

Another important aspect that must be considered is the deceleration phase. The spacecraft will be traveling at very high velocities and will not be bound to the solar system when it approaches Mars, and so it will be necessary to decrease the spacecraft's speed to reach the Martian surface without catastrophic damage. Current deceleration methods involve aerobraking \cite{Smith2005}. However, this method may not be viable at such high speeds. Aerocapture methods would allow the spacecraft to use aerodynamic drag force from a single pass through the Martian atmosphere to decelerate and achieve direct entry. Aerocapture methods can also be used for inserting a spacecraft into orbit from hyperbolic trajectories. Overall, aerocapture methods would be a good option in theory \cite{Isoletta2021}. However, these methods are again unlikely to successfully decelerate the spacecraft due to the large speeds. A futuristic proposal for deceleration involves the use of magnetic and/or electric sails that deflect trajectories of incoming ions to reduce the spacecraft speed \cite{Sharma2020}. Instead of using aerocapture methods or magnetic/electric sails, an optimal deceleration method may be to install another ground-based laser array on the surface of Mars. This would allow another laser array on Mars to decelerate the incoming spacecraft so that it can enter the entry, decent, and landing phase. However, orbital geometry limits deceleration potential, implying that payloads would need to be robust to large deceleration and impact g-loads. Additionally, it would be necessary to develop enough infrastructure on Mars to enable the installation of such a large, high power laser array, which may be very challenging to do in the near future. In addition, this would require the light sail to remain functional over the journey, or for a new sail to be deployed. At the current payload mass of 5 kg and a 5 kg solar sail, a new sail deployment would exhaust the payload capability. If the spacecraft is not decelerated, the spacecraft must have very advanced shielding that will allow it to directly enter and descend through the atmosphere and impact the Martian surface without a lot of damage. This would be challenging to accomplish due to the mass constraints of the spacecraft. During the boost, coast, and deceleration phases of the mission, the spacecraft will likely also need to perform trajectory correction maneuvers to ensure that it is on the proper trajectory. Therefore, it would likely be necessary for the spacecraft to carry lightweight microthrusters, such as miniaturized, lightweight chemical or electric thrusters found on CubeSats, on its mission to perform coarse corrections. However, given the large spacecraft velocities, trajectory correction maneuvers may also require large $\Delta v$ values, meaning that more powerful thrusters may be necessary. Specific $\Delta v$ requirements for trajectory correction maneuvers were not modeled in this work.

Lastly, geopolitical considerations of the project must be addressed as well. If a large laser array is used on Earth, there is a possibility that it could be used for targeting Earth-orbiting satellites. The lasers must also travel through Earth's atmosphere to propel the spacecraft in orbit, which is also a challenge due to potential atmospheric losses. Additionally, the lasers may interfere with or damage aerial and orbital vehicles, such as airplanes and satellites in low Earth orbit. To avoid both these issues, the laser array could instead be placed in Earth's orbit. However, this can be more challenging and expensive. Another concern is that the Earth itself could potentially be targeted by the lasers from orbit. 

\subsection{Payload vs Transit Time Analysis}
\begin{figure*} 
    \centering
    \includegraphics[width=5in]{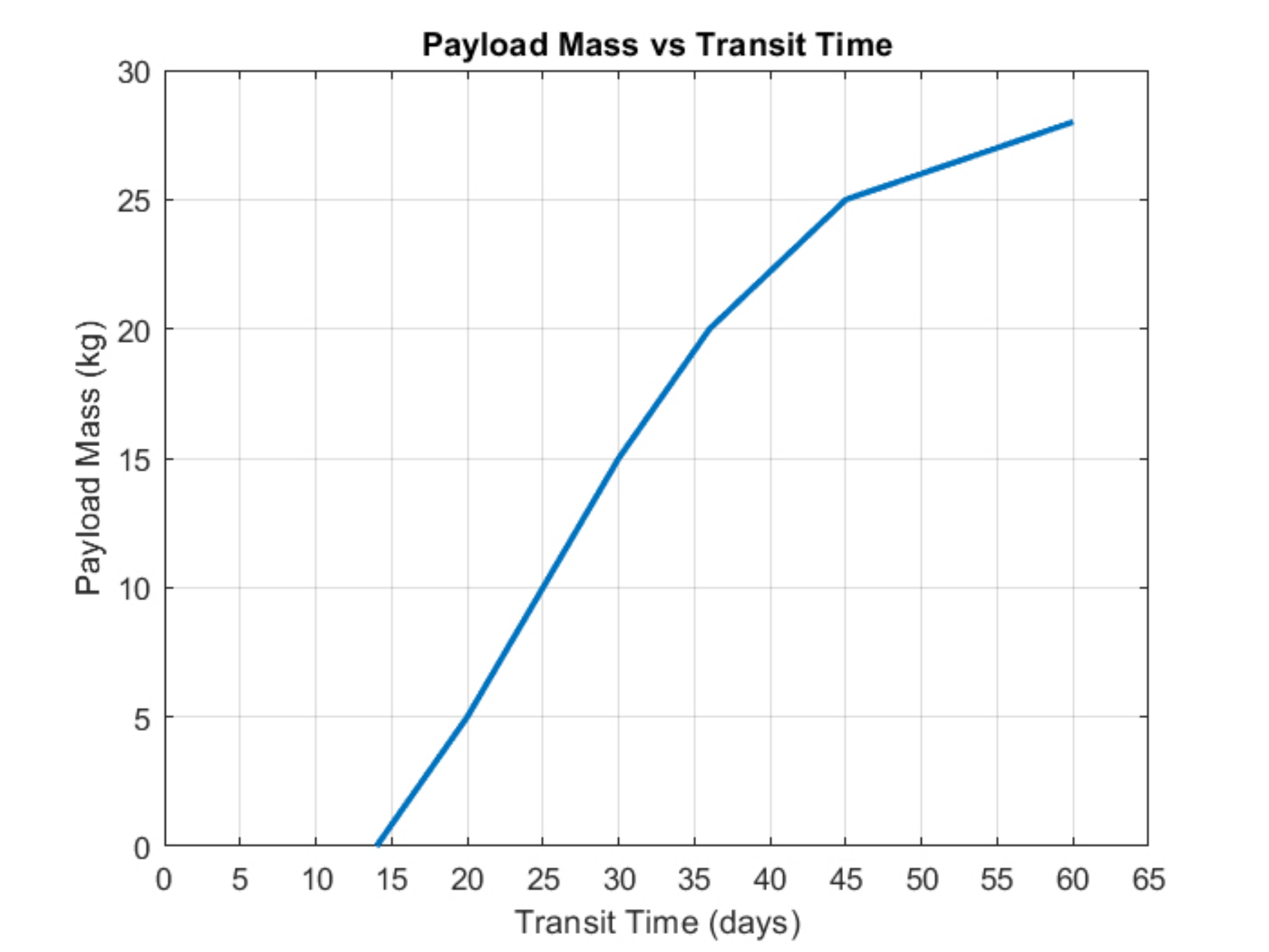}
    \caption{\bf{\label{fig:Payload vs Transit Time} Payload Mass vs Transit Time when Earth and Mars are Closest Together}}
\end{figure*}

In this paper a payload mass of 5 kg was assumed thus far. However, it is crucial to analyze the effect that increasing payload has on the transit time to Mars. When the payload is increased while keeping all other parameters constant (including optical power) the spacecraft's achievable velocity decreases. As a result, the required transit time increases. The relationship between payload mass and transit time when the spacecraft departs on June 5, 2031, the day that Mars and Earth are closest together, is shown in Figure ~\ref{fig:Payload vs Transit Time}.

By observing Figure ~\ref{fig:Payload vs Transit Time}, it can be seen that the transit time tends to increase as payload mass increases, and the relationship can be observed to initially be linear. However, as payload mass becomes larger, the relationship becomes nonlinear. This occurs after a payload mass of 20 kg. Therefore, a larger increase in transit time is required to carry more payload after the 20 kg threshold. Additionally, the maximum payload mass that can be taken in this case is 28 kg. This is because there is a minimum required departure velocity value that must be met for the mission to be possible. In this case, a minimum velocity of around 25.6 km/s is required to enable the missions given the specific departure date, optical power, and the other input characteristics. The payload mass is highest for this minimum departure velocity, which corresponds to a transit time of 60 days. For larger transit times, the departure velocity required would actually increase to more than 25.6 km/s, meaning that the payload mass would actually decrease after this point. Any departure velocity value lower than this minimum value would be infeasible, meaning that the mission would not be possible. In order to increase the payload capabilities, it is necessary to increase the optical power and change other spacecraft characteristics, such as the sail mass. 

\subsection{Boost Phase Orbit Analysis}
As discussed earlier, the boost phase requires the laser array to continuously aim lasers at the spacecraft for long periods of time. However, this is not easy to accomplish due to the visibility limitations of the spacecraft while in Earth’s orbit from the location of the laser array on Earth. Similar to a satellite communicating with a ground station, the spacecraft must be visible from the laser array on Earth in order to be boosted by the lasers. Installing a single laser array of the required magnitude on Earth would be optimal; multiple laser arrays would increase installment and operating costs and complexity. Therefore, an optimal initial orbit must be chosen for the spacecraft to coast in during the boost phase. 

To allow the spacecraft to be continuously visible from a location on the Earth’s surface, the spacecraft may be inserted into a geostationary orbit after launch. This allows the spacecraft to be continuously visible from a specific location on Earth with no boost time constraints, and so the vehicle can be continuously boosted to the target velocity. However, a geostationary orbit is not ideal because it is a high-altitude orbit 35,786 km above the Earth’s equator. Therefore, the lasers may affect the multitude of satellites in low Earth orbit and medium Earth orbit. It would be challenging to ensure that the lasers would not affect other satellites in lower Earth orbits.

In order to avoid the issues with the geostationary orbit, the laser array could be moved from the Earth's surface and be installed on the far side of the Moon. As a result, laser interference with aerial traffic and orbital traffic could be avoided since the Moon has no aerial traffic, and orbital traffic is very low. Additionally, critical geopolitical issues could be avoided as well. If the laser array is placed on the far side of the Moon, there will be no way for it to potentially be aimed at the Earth’s surface or at Earth-orbiting satellites. Therefore, the two important geopolitical issues would be resolved, and safe use of the laser system would be ensured. The Moon does not have an atmosphere, and so an added benefit is that there will be no atmospheric losses for the lasers as there would be on Earth. Specifically, the spacecraft could be inserted into a halo orbit around the Earth-Moon L2 point. This orbit ensures that the spacecraft is continuously visible to the laser array from the far side of the Moon, and so it can be continuously boosted to the target velocity. Due to the immense interest in setting up a lunar base and the Artemis missions, it is very likely that proper infrastructure will be developed on the Moon in the near future. Additionally, target velocity requirements from the Moon are very similar to the velocities required from Earth. Therefore, setting up a laser array on the Moon would be a realistic possibility. Overall, setting up the laser array on the Moon and placing the spacecraft in a halo orbit around the Earth-Moon L2 point may be an interesting boost phase orbit option for 20-day rapid transit missions to Mars. 

\subsection{Cost Analysis}
The cost of the 20-day rapid missions is another important factor to consider. By assuming a nominal electrical energy cost of \$0.1/kW-hr, the cost to produce the required amount of energy for the missions was computed given a specific departure date and the corresponding optical power requirement. The results are shown in Table ~\ref{tab:cost}. It can be observed that the electrical energy cost varies from \$ 6 million to \$ 30 million for the 27 month 2030-2032 time period considered. Taking launch costs at 20 $k/kg$, a 10 kg spacecraft launch would be \$200k, a small fraction of the total cost; therefore, launch costs are ignored in our analysis.

The lower costs occur when lower power is required, meaning a lower departure velocity is necessary. This corresponds to the time around when Mars is at opposition. The highest cost would be incurred around the time of solar conjunction, when Mars is farther from Earth, requiring higher velocity and power. Current missions to Mars range in cost; Perseverance was the third most expensive Mars mission and cost a total of \$ 2.725 billion, which equates to \$2.1 million/kg for the overall mission. However, the 20-day rapid transit mission would have a size comparable to a 2.5U CubeSat, meaning it would transport a very limited amount of payload and cargo, constrained to less than 5 kg. The electrical energy costs are relatively modest, and the development cost of a 2.5U CubeSat is relatively low due to the extensive expertise and commercial market for the area. Table ~\ref{tab:cost} shows the total electrical energy cost and the cost per kg of mass for the rapid transit missions. It can be observed that the \$M/kg ranges from 1.22 to 5.92 \$M/kg. The lower end of these values are comparable to the Mars Perseverance overall \$M/kg value, except these costs exclude the Mars Perseverance mission development and launch costs. Light sails are a proven technology, but will likely remain expensive in the near future. Sail costs may be variable depending on required improvements to the sail to allow exposure to the large amount of power from the lasers. Energy storage costs may be substantially high due to the high amount of power being used. Lastly, the deceleration phase will likely need to use novel and expensive technologies and methods. The entry, descent, and landing phase will cost a substantial amount but would vary based on technologies used. Advanced methods such as the sky crane would not be possible due to the mass limitations, so a complex, mass-constrained method must be developed for this phase. The novel concepts for deceleration and entry, descent, and landing would almost certainly be expensive, at least for the first time they are implemented. Therefore, it may be most beneficial to solely use this mission concept for urgent and emergency missions where very rapid transit is absolutely necessary.   

\begin{table}[]
\renewcommand{\arraystretch}{1.3}
\centering
\caption{\label{tab:cost}Rapid Transit Mission Cost Estimates}
\begin{tabular}{|l|l|l|l|} \hline
\textbf{\begin{tabular}[c]{@{}l@{}}Departure \\ Date\end{tabular}} & \textbf{\begin{tabular}[c]{@{}l@{}}Optical Power \\ (GW)\end{tabular}} & \textbf{\begin{tabular}[c]{@{}l@{}}Electrical  \\ Energy Cost \\ (Millions)\end{tabular}} & \textbf{\begin{tabular}[c]{@{}l@{}}\$/kg \\ (Millions   \\ /kg)\end{tabular}} \\\hline
5/25/2030 & 11.732 & 28.3 & 5.66 \\\hline
6/25/2030 & 12.283 & 29.0 & 5.80 \\\hline
9/5/2030 & 11.732 & 28.3 & 5.66 \\\hline
12/1/2030 & 7.430 & 22.5 & 4.50 \\\hline
2/1/2031 & 3.730 & 16.0 & 3.20 \\\hline
4/4/2031 & 1.261 & 9.29 & 1.86 \\\hline
5/4/2031 & 0.719 & 7.01 & 1.40 \\\hline
6/5/2031 & 0.546 & 6.11 & 1.22 \\\hline
9/5/2031 & 1.528 & 10.2 & 2.04 \\\hline
12/1/2031 & 3.791 & 16.1 & 3.22 \\\hline
3/1/2032 & 7.693 & 22.9 & 4.58 \\\hline
6/11/2032 & 12.171 & 28.9 & 5.78 \\\hline
7/11/2032 & 12.846 & 29.6 & 5.92 \\\hline
\end{tabular}
\end{table}

\subsection{Applications of the Mission Concept}
\begin{figure*} 
    \centering
    \includegraphics[width=5in]{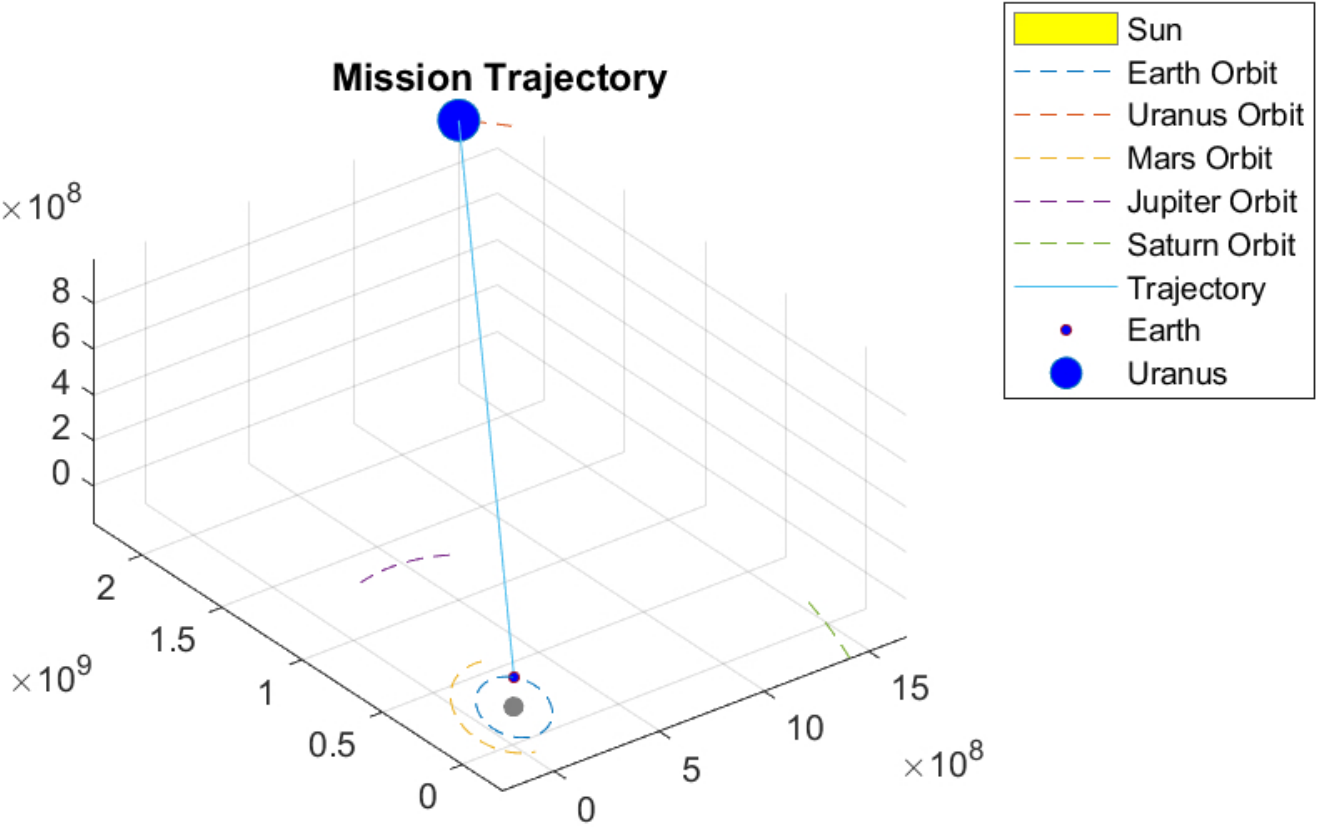}
    \caption{\bf{\label{fig:EarthUranus}1-Year Trajectory with Departure on Uranus Opposition}}
\end{figure*}

The 20-day rapid transit missions to Mars can serve as precursor missions to more complex, distant missions and destinations with varying goals. Such missions may be used for expedited transit, exploration, and study of lesser studied  bodies in the solar system such as Uranus, Neptune, and Pluto. Potential missions to asteroids and comets would also have a dramatically decreased transit time. Since deceleration and entry, descent, and landing are challenging, critical planetary science missions involving flybys and impactors are ideal. Flyby missions and impactor missions to celestial bodies, such as Triton, Uranus, and Europa, can be accomplished by using this mission concept. Impactor missions with strong and robust structural protection systems would allow lander missions as well. An interesting impactor mission concept would allow the spacecraft to test the creation of liquid brines upon impact. Another interesting mission concept could test the concept of Panspermia, which revolves around the theory that life originated in space or on another planetetary body (e.g., Mars) but was deposited elsewhere via meteoritic transfer.  Lasers may be used for planetary defense against asteroids. This would involve aiming the lasers at asteroids to deflect their trajectories away from Earth. Missions to interstellar objects such as ${'}$Oumuamua and interstellar destinations such as Alpha Centauri would be possible in a matter of decades, as opposed to the thousands of years the transit would take if current technologies were used \cite{Litchford2020}. Figure ~\ref{fig:EarthUranus} shows a 1 year Earth-Uranus rapid transit mission with a 5 kg payload when Uranus is at opposition. The Voyager missions took 8.5 years to reach Uranus, but this mission would take just 11.8 percent of that time. A 1 year mission to Uranus is possible if the departure velocity is 96.06 km/s, meaning 2.331 GW of power would be necessary. Technologies enabling such rapid transit missions can be developed in the next several decades and be applied to a variety of innovative and valuable missions that will allow humanity to pursue valuable knowledge regarding the universe, defend Earth, and also become an interplanetary civilization.

\section{Conclusions}
Light sails can be used in conjunction with an Earth-based array of lasers as a means of propulsion to enable rapid 20-day transit missions from Earth to Mars. The time frame between April 25, 2030-August 10, 2030 provides a window during which two solar conjunctions and one Martian opposition occurs. It is found that 20-day rapid transit missions to Mars are possible during the majority of times within this time frame. The 20-day trajectories are all hyperbolic trajectories in which the spacecraft are not bound to the solar system. During this time frame, the only times when 20-day missions are unlikely to be feasible are on and around solar conjunction, when the spacecraft's trajectory is in very close proximity to the Sun. This issue can be resolved by extending the spacecraft transit duration for departure around solar conjunction. The optical power required to enable these 20-day missions within the chosen time frame ranges from 0.546 GW to 13.075 GW and are based on the spacecraft velocities required for the various departure dates. This means that the laser array must be able to provide a range of power. The boost phase should be minimized to continuously boost the spacecraft from Earth, but this would require very high optical power values. Optimal departure orbits from the Earth or Moon may be chosen to allow for the continuous boost of the spacecraft without a time constraint. To avoid high power values, less payload mass would need to be used. Additionally, the lasers would need to change their orientations in order to continuously track the spacecraft or several laser arrays would be required. Deceleration can be done by using an alternate laser array on Mars, while electric and/or magnetic sails and advanced aerocapture methods should be considered as well. Given proper funding for further research, technologies enabling 20-day missions to Mars can be developed within the next few decades. Such missions to Mars will facilitate the exploration of Mars and the creation of Martian infrastructure and can provide key supplies required for a human settlement in a timely manner. These missions will allow extensive technology development such as more powerful lasers, stronger and more robust materials, the miniaturization of key technologies, and cheaper mass production of products usable in space and back on Earth. Lastly, these missions would support the technological advancement of humanity, facilitate the pursuit of knowledge of the universe, and serve as precursor missions to more distant solar system objects and interstellar missions.



\acknowledgments
Funded by faculty startup funds from the Georgia Institute of Technology to C.E.C.

\bibliographystyle{IEEEtran}
\newcommand\BIBentryALTinterwordstretchfactor{1}
\bibliography{Bibliography}

\begin{thebibliography}{10}
\providecommand{\url}[1]{#1}
\csname url@samestyle\endcsname
\providecommand{\newblock}{\relax}
\providecommand{\bibinfo}[2]{#2}
\providecommand{\BIBentrySTDinterwordspacing}{\spaceskip=0pt\relax}
\providecommand{\BIBentryALTinterwordstretchfactor}{4}
\providecommand{\BIBentryALTinterwordspacing}{\spaceskip=\fontdimen2\font plus
\BIBentryALTinterwordstretchfactor\fontdimen3\font minus
  \fontdimen4\font\relax}
\providecommand{\BIBforeignlanguage}[2]{{%
\expandafter\ifx\csname l@#1\endcsname\relax
\typeout{** WARNING: IEEEtran.bst: No hyphenation pattern has been}%
\typeout{** loaded for the language `#1'. Using the pattern for}%
\typeout{** the default language instead.}%
\else
\language=\csname l@#1\endcsname
\fi
#2}}
\providecommand{\BIBdecl}{\relax}
\BIBdecl

\bibitem{Litchford2020}
R.~J. Litchford and J.~A. Sheehy, ``Prospects for interstellar propulsion,'' in
  \emph{Annual AAS Guidance, Navigation and Control Conference}.\hskip 1em plus
  0.5em minus 0.4em\relax AAS, 2020, pp. AAS--20--068, 1--12.

\bibitem{PARKIN2018370}
\BIBentryALTinterwordspacing
K.~L. Parkin, ``The breakthrough starshot system model,'' \emph{Acta
  Astronautica}, vol. 152, pp. 370--384, 2018. [Online]. Available:
  \url{https://doi.org/10.1016/j.actaastro.2018.08.035}
\BIBentrySTDinterwordspacing

\bibitem{Tanaka2020}
\BIBentryALTinterwordspacing
K.~A. Tanaka, K.~M. Spohr, D.~L. Balabanski, S.~Balascuta, L.~Capponi, M.~O.
  Cernaianu, M.~Cuciuc, A.~Cucoanes, I.~Dancus, A.~Dhal, B.~Diaconescu,
  D.~Doria, P.~Ghenuche, D.~G. Ghita, S.~Kisyov, V.~Nastasa, J.~F. Ong,
  F.~Rotaru, D.~Sangwan, P.-A. Söderström, D.~Stutman, G.~Suliman,
  O.~Tesileanu, L.~Tudor, N.~Tsoneva, C.~A. Ur, D.~Ursescu, and N.~V. Zamfir,
  ``Current status and highlights of the eli-np research program,''
  \emph{Matter and Radiation at Extremes}, vol.~5, no.~2, p. 024402, 2020.
  [Online]. Available: \url{https://doi.org/10.1063/1.5093535}
\BIBentrySTDinterwordspacing

\bibitem{ACTON199665}
\BIBentryALTinterwordspacing
C.~H. Acton, ``Ancillary data services of nasa's navigation and ancillary
  information facility,'' \emph{Planetary and Space Science}, vol.~44, no.~1,
  pp. 65--70, 1996, planetary data system. [Online]. Available:
  \url{https://doi.org/10.1016/0032-0633(95)00107-7}
\BIBentrySTDinterwordspacing

\bibitem{ACTON20189}
\BIBentryALTinterwordspacing
C.~Acton, N.~Bachman, B.~Semenov, and E.~Wright, ``A look towards the future in
  the handling of space science mission geometry,'' \emph{Planetary and Space
  Science}, vol. 150, pp. 9--12, 2018, enabling Open and Interoperable Access
  to Planetary Science and Heliophysics Databases and Tools. [Online].
  Available: \url{https://doi.org/10.1016/0032-0633(95)00107-7}
\BIBentrySTDinterwordspacing

\bibitem{Blanchard1969}
E.~L. R.C.~Blanchard, ``A unified form of lambert's theorem,'' pp.
  NASA--TN--D--5368, 1--18, 1969.

\bibitem{Gooding1990}
\BIBentryALTinterwordspacing
R.~Gooding, ``A procedure for the solution of lambert's orbital boundary-value
  problem,'' \emph{Celestial Mechanics and Dynamical Astronomy}, vol.~48, pp.
  145--165, 1990. [Online]. Available: \url{https://doi.org/10.1007/BF00049511}
\BIBentrySTDinterwordspacing

\bibitem{rodyo}
\BIBentryALTinterwordspacing
Rodyo, ``Release v1.4: Create readme.md · rodyo/fex-lambert.'' [Online].
  Available: \url{https://github.com/rodyo/FEX-Lambert/releases/tag/v1.4}
\BIBentrySTDinterwordspacing

\bibitem{Lubin2016}
\BIBentryALTinterwordspacing
P.~Lubin, ``A roadmap to interstellar flight,'' 2016. [Online]. Available:
  \url{https://arxiv.org/abs/1604.01356}
\BIBentrySTDinterwordspacing

\bibitem{Tung2021}
\BIBentryALTinterwordspacing
H.-T. Tung and A.~Davoyan, ``Light-sail photonic design for fast-transit earth
  orbital maneuvering and interplanetary flight,'' 2021. [Online]. Available:
  \url{https://arxiv.org/abs/2107.09121}
\BIBentrySTDinterwordspacing

\bibitem{DUPLAY2022143}
\BIBentryALTinterwordspacing
E.~Duplay, Z.~F. Bao, S.~{Rodriguez Rosero}, A.~Sinha, and A.~Higgins, ``Design
  of a rapid transit to mars mission using laser-thermal propulsion,''
  \emph{Acta Astronautica}, vol. 192, pp. 143--156, 2022. [Online]. Available:
  \url{https://doi.org/10.1016/j.actaastro.2021.11.032}
\BIBentrySTDinterwordspacing

\bibitem{Venzmer2018}
\BIBentryALTinterwordspacing
{Venzmer, M. S.} and {Bothmer, V.}, ``Solar-wind predictions for the parker
  solar probe orbit - near-sun extrapolations derived from an empirical
  solar-wind model based on helios and omni observations,'' \emph{A\&A}, vol.
  611, p. A36, 2018. [Online]. Available:
  \url{https://doi.org/10.1051/0004-6361/201731831}
\BIBentrySTDinterwordspacing

\bibitem{Smith2005}
\BIBentryALTinterwordspacing
J.~C. Smith and J.~L. Bell, ``2001 mars odyssey aerobraking,'' \emph{Journal of
  Spacecraft and Rockets}, vol.~42, no.~3, pp. 406--415, 2005. [Online].
  Available: \url{https://doi.org/10.2514/1.15213}
\BIBentrySTDinterwordspacing

\bibitem{Isoletta2021}
\BIBentryALTinterwordspacing
G.~Isoletta, M.~Grassi, E.~Fantino, D.~de~la Torre~Sangr\`{a}, and
  J.~Pel\'{a}ez, ``Feasibility study of aerocapture at mars with an innovative
  deployable heat shield,'' \emph{Journal of Spacecraft and Rockets}, vol.~58,
  no.~6, pp. 1752--1761, 2021. [Online]. Available:
  \url{https://doi.org/10.2514/1.A35016}
\BIBentrySTDinterwordspacing

\bibitem{Sharma2020}
C.~W. Kush K.~Sharma and A.~M. Hein, ``Feasibility assessment of deceleration
  technologies for interstellar probes,'' in \emph{International Astronautical
  Congress 20 - CyberSpace Edition}.\hskip 1em plus 0.5em minus 0.4em\relax
  International Astronautical Federation, 2020, pp. 1--9.

\end{thebibliography}

\thebiography
\begin{biographywithpic}
{Madhukarthik Mohanalingam}{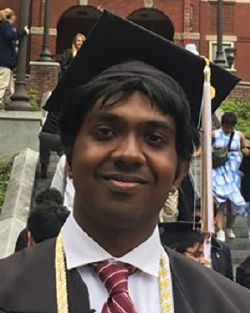} is a Masters student at the Georgia Institute of Technology. He received his Bachelor of Science in Aerospace Engineering in 2022 from the Georgia Institute of Technology. He is a student researcher at the Georgia Tech Space Systems Design Laboratory.
\\
\end{biographywithpic} 

\begin{biographywithpic}
{Christopher E. Carr}{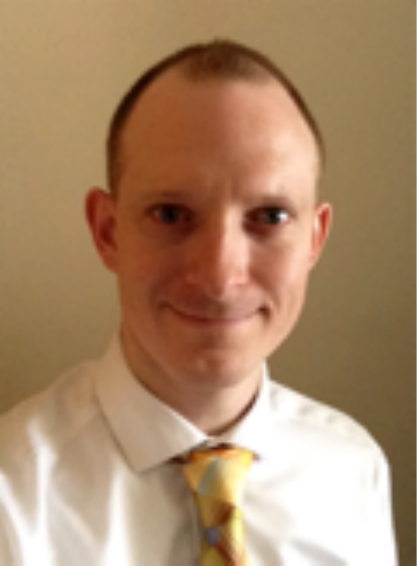} received his B.S. degree in Aero/Astro and Electrical Engineering in 1999, his Masters degree in Aero/Astro in 2001, and his Sc.D. degree in Medical Physics in 2005, all from MIT. He is an Assistant Professor in the Daniel Guggenheim School of Aerospace Engineering and the School of Earth and Atmospheric Sciences at the Georgia Institute of Technology. He is broadly interested in searching for and expanding the presence of life beyond Earth.
\end{biographywithpic}

\end{document}